# An open-source Modular Online Psychophysics Platform (MOPP)

Yuval Samoilov-Kats, Matan Noach, Noam Beer, Yuval Efrati and Adam Zaidel

Gonda Multidisciplinary Brain Research Center, Bar Ilan University, Ramat Gan, Israel

**Financial disclosures:** None to report

**Funding sources:** This research was supported by grants from the Israel Science Foundation (ISF, grant No. 1291/20) and the Data Science Institute at Bar-Ilan University to AZ.

**Abstract**

In recent years, there is a growing need and opportunity to use online platforms for psychophysics research. Online experiments make it possible to evaluate large and diverse populations remotely and quickly, complementing laboratory-based research. However, developing and running online psychophysics experiments poses several challenges: i) a high barrier-to-entry for researchers who often need to learn complex code-based platforms, ii) an uncontrolled experimental environment, and iii) questionable credibility of the participants. Here, we introduce an open-source Modular Online Psychophysics Platform (*MOPP*) to address these challenges. Through MOPP's simple web-based interface, researchers can build modular experiments, share them with others, and copy or modify tasks from each other's environments. MOPP provides built-in features to calibrate for viewing distance and to measure visual acuity. It also includes email-based and IP-based authentication, and 'reCAPTCHA' verification. We developed five example psychophysics tasks, that come preloaded in the environment, and ran a pilot experiment which was hosted on the AWS (Amazon Web Services) cloud. Pilot data collected for these tasks yielded similar results to those reported in laboratory settings. MOPP can thus help researchers collect large psychophysics datasets online, with reduced turnaround time, and in a standardized manner.

# Introduction

Psychophysics experiments that require the physical presence of participants in a laboratory suffer from limitations in sample size and recruitment options. For example, they often underrepresent participants with accessibility and mobility limitations and overrepresent student populations (Druckman & Kam, 2009; Sears, 1986). Moreover, data acquisition can be affected or even halted, by external circumstances, as exemplified by the COVID-19 pandemic shutdowns. Running experiments online can overcome these constraints, since participants can take part outside the laboratory, in the convenience of their own homes.

Psychophysics experiments vary broadly in their structure, design, and the type of stimuli they use, depending on the scientific objectives. Hence, tools for conducting psychophysics experiments online need to allow for flexible customization. However, there is no unified platform that singlehandedly enables this. Instead, researchers are generally required to integrate multiple platforms, each with different functionalities. Specifically, at least two platforms are needed: one for building the experiment and another for hosting it. This can be a complex, expensive, and skill-dependent process, considerably raising the barrier-to-entry for some researchers and students.

The scientific community has developed many valuable and open-access tools for this purpose (de Leeuw, 2015; Lago, 2021; Lange et al., 2015; Onnela et al., 2021; Peirce et al., 2019; Schwarzbach, 2011; Torous et al., 2016; Woods et al., 2015); see Grootswagers, 2020 for an overview). Table 1 lists several popular alternatives for building psychophysics experiments (left half of the table) and their corresponding platforms for hosting experiments (right half of the table; although other combinations are also possible). However, building custom experiments with these platforms often requires knowledge of their respective programming languages and/or familiarity with dedicated software applications. Moreover, additional programming skills are required to configure the servers, whether the experiment is hosted on a local server or hosted using cloud services.

**Table 1 – Popular platforms for building and hosting psychophysical experiments online.**

| Building experiments | | Hosting experiments | |
|---|---|---|---|
| **Platform** | **Programming language** | **Platform** | **Website address** |
| PsychoPy (Peirce et al., 2019) | Python | Pavlovia (Bridges et al., 2020) | https://pavlovia.org |
| JsPsych (de Leeuw, 2015) | JavaScript | JATOS (Lange et al., 2015) e.g., using MindProbe | https://jatos.org https://mindprobe.eu/ |
| Psychtoolbox (Brainard, 1997) | MATLAB | LabMaestro Pack & Go | https://vpixx.com/products/labmaestro-packngo/ |

Another critical challenge in conducting psychophysics experiments online is the lack of control over the experimental environment, compared to laboratory-based experiments. If not taken into account, confounding factors, such as visual acuity and variations in viewing distance, add noise to the experimental measures obtained. Additionally, many current online platforms do not validate the credibility of participants adequately (Chandler et al., 2014; Chmielewski & Kucker, 2020; Kennedy et al., 2020; Paolacci & Chandler, 2014; Peer et al., 2022). This reduces data quality and reliability. Thus, a unified and user-friendly platform for online experiments with improved experimental control and participant validation, would provide a valuable contribution.

This paper presents an open-source Modular Online Psychophysics Platform (*MOPP*). MOPP provides a simple web-based interface to build, manage, and launch online experiments. Additionally, it has integrated tools to calibrate for viewing distance, measure visual acuity, and confirm participants' credibility. We collected pilot data for five example tasks that come preloaded in the MOPP environment. The results were comparable to those reported when run in laboratory settings. In the following three sections, we provide an overview of MOPP's architecture (Section A), describe a typical workflow (Section B), and present pilot data gathered with MOPP (Section C).

## A. MOPP architecture

The MOPP architecture consists of three integrated components (Figure 1a). i) The server-side component. This contains the code for running experiments. It controls the progression of the experiments, sends stimulus details to the participant's web browser (on the client side), and receives information in return (e.g., responses). It runs in 'Node.js' – an open-source JavaScript (JS) server environment. ii) The database component. This enables the efficient management and storage of the experimental data. It runs on the server and interacts with the server-side component for data exchange. This was implemented using 'MongoDB', an open-source database management program. iii) The client-side component. This provides a web-based interface for human interaction with MOPP. It provides access either as a participant (to perform the experiment) or as a researcher (for experiment management). On the participant's web browser, it displays experiment-related content, including text, stimuli, and input fields. On the researcher side, it displays a web interface for building, launching, and managing experiments. The client-side component was developed using 'React.js', an open-source JS library for user interfaces.

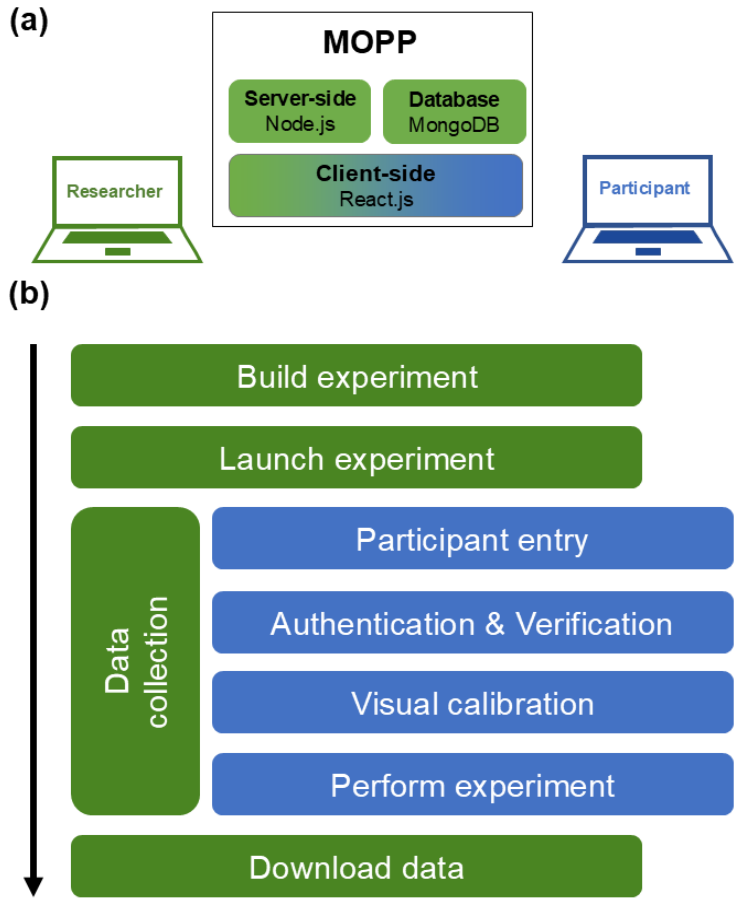

**Figure 1: MOPP overview.** (a) MOPP consists of three main components: (i) the server-side component - 'Node.js', (ii) the database component - 'MongoDB', and (iii) the client-side component - 'React.js'. The researcher (green) interacts with all three components, while the participant (blue) only interacts with the client side. (b) A typical workflow on the platform. The researcher first builds an experiment (by adding experimental tasks and defining their details) and then launches it. During data collection, participants enter the experiment and undergo email and IP-address authentication, and reCAPTCHA verification. Then, they undergo visual calibration tests to calibrate for viewing distance and to measure visual acuity (per participant). After completing these steps, participants perform the experiment. Finally, the researcher downloads the data.

**Code availability and installation process**

MOPP is an open-source project, and the code is publicly available at: https://gitlab.com/mopp.proj/mopp-project/. Before running experiments with MOPP, the researcher first needs to complete the setup process by following the step-by-step instructions in the *MOPP guide* (available on the project's Gitlab page). The MOPP guide explains the process of hosting an experiment using the AWS cloud (this requires creating an AWS account). We chose AWS because of its popularity and built-in services. For example, it offers adjustable computational power and storage capacity according to the experimental requirements. MOPP can also be hosted on other cloud servers (e.g., Microsoft Azure or Google Cloud) or on a local server, however, we only cover AWS hosting in the scope of this paper.

## B. MOPP workflow

Figure 1b presents a typical workflow on the platform. It consists of four main steps (green boxes). The first step begins with the researcher, who builds an experiment according to their scientific objectives. This includes adding the relevant tasks and defining the details of the stimuli. In the second step, the researcher launches the experiment and shares a dedicated link with the participants, by which they access the experiment. The third step, data collection, involves the participants (blue boxes). During this step, the participants enter the experiment via a web browser from their personal computers (PCs). The participants then undergo several tests, including authentication and verification of the participants' credibility, and visual calibration tests to account for viewing distance and to measure visual acuity. Following these tests, the participants perform the actual experiment. Finally, in the last step, the researcher downloads the acquired data via MOPP. This workflow is further described in detail below.

**1. Build experiment**

The researcher builds an experiment through the *researcher portal* (Figure 2a), which is accessed via a computer web browser. The researcher portal displays a summary table of the stored experiments. The experiment name, number of tasks within each experiment, how many participants have accessed the experiment, and how many of those have completed it, are presented in the table (some details are omitted in Fig. 2a, for simplicity). From the summary table, the researcher can open the *experiment page* (Figure 2b; each experiment has its own separate page). There, the researcher can build, edit, and launch the experiment.

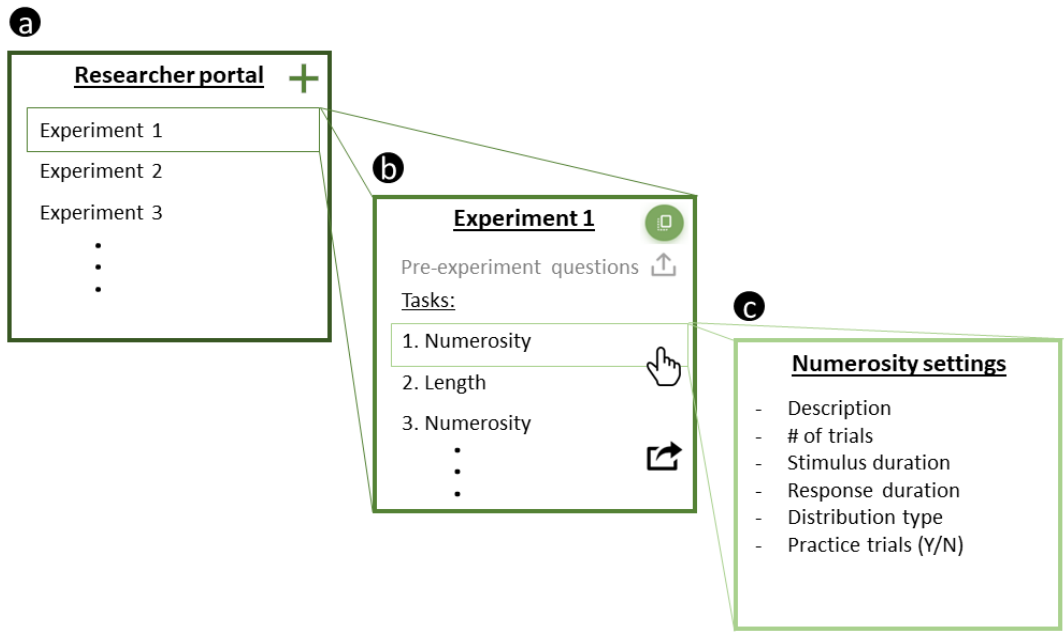

**Figure 2: Illustration of the web-based interface for managing, building, and launching experiments.** (a) The researcher portal displays a list of saved experiments with key information. On the researcher portal, the researcher can create a new experiment (using the ✚ button). Clicking on the experiment name (e.g., 'Experiment 1') opens the experiment page. (b) On the experiment page, the researcher can clone a previous experiment (using the ⊙ button), define the order of the tasks (drag-and-drop using the 🖑 cursor), and launch the experiment (using the ↪ button). The pre-experiment questions can be edited (using the ⬆ button). Clicking on a task name, e.g., 'Numerosity', opens the task page (c), where task-specific settings can be modified.

To get started, the researcher can choose an experiment from the preloaded list of sample experiments, or create a new one (plus button in Figure 2a). The researcher can copy an experiment with all its settings using the *clone feature* (green button in Figure 2b) and then edit its details. The clone feature enables the researcher to copy and edit an experiment under a new name without changing the original.

**1.1 . Pre-experiment questions**

Many psychophysics experiments begin with a set of self-reported questions. These typically ask participants to give informed consent, to provide information regarding demographics (e.g., age, gender), to report whether they wear glasses, and whether they suffer from medical conditions (neurological, psychiatric, etc.). By default, MOPP adds these questions to the beginning of each experiment.

The researcher can replace these default questions by uploading a custom set of questions (light gray button in Figure 2b). For instance, if a crowdsourcing platform is used to recruit participants (e.g., Prolific or MTurk), a question can be added to obtain the participants' platform-user-IDs for payment approval. To replace the default questions, the researcher must upload a JavaScript Object Notation (JSON) file containing the new set of questions. Several websites can be used to create and download JSON files, such as Survey.io ([https://surveyjs.io/create-survey](https://surveyjs.io/create-survey)) or Qualtrics ([https://www.qualtrics.com](https://www.qualtrics.com)). Thus, no programming skills are required to upload a new set of questions.

## 1.2 Preloaded tasks

In order to demonstrate MOPP functionality, five example tasks were developed and loaded to MOPP, as described in Table 2.

**Table 2 – Description of the five preloaded tasks**

| Task name | Stimulus | Task | Conditions |
| --- | --- | --- | --- |
| *Length* | The stimulus is a series of connected underbars (the '_' symbol) that form a solid line. On the task page (Fig. 2c) the researcher can define the distribution (Uniform, Gaussian, or Bimodal) and its parameters, for the number of underbars per experiment (fixed across participants). A schematic of the stimulus is presented in Figure 3. | Participants are asked to estimate, without counting, the length of the line (i.e., to estimate the number of underbars). | No predefined conditions. |
| *Numerosity* | The stimulus is an array of black and white dots randomly distributed on a gray background. On the task page (Fig 2c), the researcher can define the distribution (Uniform, Gaussian, or Bimodal) and its parameters, for the number of dots. A schematic of the stimulus is presented in Figure 4. | Participants are asked to estimate, without counting, the total number of dots (irrespective of color). | No predefined conditions. |
| *Biological motion* | The stimulus is a set of white moving dots on a black background. The stimulus set either represents: i) biological motion (*BM*) - that resembles natural human movement or, ii) random motion (*RM*) - a scrambled version of the BM, moving in an unspecific pattern that does not resemble BM. Stimulus schematics are presented in Figure 5.<br>Stimuli were adapted from (Troje, 2002;using the online simulator at: [https://www.biomotionlab.ca/html5-bml-walker/](https://www.biomotionlab.ca/html5-bml-walker/)). | Participants are asked to discriminate between BM and RM stimuli. | The researcher can choose from three predefined conditions, with different numbers of dots: 1) the BM and RM stimuli contain 15 moving dots (easier to detect). 2) both types of stimuli contain 10 moving dots. |

| | | | 3) both types of stimuli contain 7 moving dots. |
|---|---|---|---|
| *Mooney face* | The stimulus is a set of black-and-white images that either contain faces or not. Faces can be oriented upright or inverted (the former are easier to detect). Images without faces (*random* image) are scrambled versions of those with faces. Stimulus schematics are presented in Figure 6. Stimuli were adapted from Schwiedrzik et al., 2018. | Participants are asked to discriminate between faces (irrespective of upright/inverted orientation) and random image stimuli. | The researcher can choose from three predefined conditions, using: 1) only upright faces and random image stimuli. 2) all three types of stimuli (i.e., upright faces, inverted faces, and random image stimuli). 3) only inverted faces and random image stimuli. |
| *Key-tapping* | This is a motor task, so there is no stimulus to present. Rather, a countdown timer is displayed, indicating the time within which the participant must complete the task. A schematic of the task is presented in Figure 7. This task was adapted from (Noyce et al., 2014), modified to test both hands. | Participants are asked to press the 's' and 'k' keys alternatingly on their keyboard within a fixed 30-second interval. | The researcher can choose from three predefined conditions: i) key tapping with both hands, ii) right hand only, or iii) left hand only. These differ only in the instructions given to participants. |

## 1.3 Adding tasks to an experiment

The researcher can add the preloaded tasks to a specific experiment using MOPP's interface, without programming, from a drop-down list (not shown in Fig. 2b for simplicity). The settings of the tasks can be modified from the *task page* (Figure 2c), where the task description, number of trials, stimulus duration, and response duration can be set.

## 1.4 Creating a new task

Beyond the preloaded tasks, if a new task is required, this must be designed and developed in advance (or copied from another user). With MOPP, all tasks should be developed in JS, a popular web development language. Researchers proficient in JS programming can develop tasks from scratch, directly in JS. Otherwise, tasks can be developed without JS expertise using the jsPsych platform (de Leeuw, 2015). With jsPsych, tasks are created from *plugins* – templates of JS code with experimental events, task components, or even complete standalone tasks. Researchers can choose from an extensive library of available plugins to create different tasks, and then load their new tasks to MOPP.

When developing a new task, the researcher should decide which task-specific settings will be modifiable from the task page (rather than having to modify the code). For instance, if a task's stimulus magnitude is obtained from a certain distribution, the researcher can set the distribution parameter(s) from the task page. Including modifiable task-specific settings during development provides flexibility to the researcher, and prevents the need for additional (re)programming.

After developing a new task, it should be loaded into MOPP. Then it can be added to experiments (similar to the preloaded tasks). The MOPP user guide demonstrates how to load a task designed in jsPsych to MOPP (section 2c of the guide). It takes the researcher step-by-step through this process with the Random Dot Kinematogram (RDK) task (Rajananda et al., 2018) as an example.

To ensure task understanding and to familiarize participants with the response format, researchers can use MOPP's practice trial feature. Practice trials help improve data quality and reduce the likelihood of excluding participants due to task misunderstanding - a common challenge in online experiments (Chandler et al., 2014; Chmielewski & Kucker, 2020; Kennedy et al., 2020; Paolacci & Chandler, 2014; Peer et al., 2022). The researcher can set predefined conditions to confirm that the participants have responded reasonably in the practice trial before continuing the experiment (e.g., that the response is correct or within a certain range). These conditions should be defined during task development. The researcher can then choose whether or not to use it in a specific experiment from the task page (Figure 2c).

2. **Launch experiment**

The researcher launches the experiment from the experiment page (Figure 2b). This generates a shareable link for participants to access the experiment. Once an experiment is launched, its settings become uneditable. If required, the researcher can clone a launched experiment which can be edited without affecting the original experiment. Before launching the experiment, the researcher must choose between one of two possible modes for running the experiment:

i) *Public* mode. This is meant for remote (online) experiments conducted outside the laboratory. In public mode, the participants first complete authentication and verification tests (described in subsection 3.2.), then they answer the pre-experiment questions (described in subsection 1.2.), followed by the visual calibration tests (described in subsection 3.3.). After these are completed, they perform the experiment.

ii) S*upervised* mode. This is meant for experiments conducted in a controlled physical setting, such as a laboratory or clinic, using the MOPP platform. In this mode, the researcher sets up and controls the physical environment. The researcher also personally verifies each participant's identity. Therefore, participants skip the authentication, verification, and visual calibration tests. Researchers interested in performing the calibration tests in this mode (to measure participants viewing distance and/or visual acuity) could activate these tests via MOPP's source code or, instead, create and load it as a task within the experiment.

## 3. Data collection

### 3.1. Participant entry

Participants enter the experiment from the web browser of their PCs – laptop or desktop. Currently, MOPP supports only PCs (not mobile devices, such as tablets or smartphones). Thus, the researcher should instruct participants to use a PC. If participants attempt to enter using a mobile device, an error page will be presented, instructing them to access the experiment using a PC. Participants should also be instructed to deactivate any advertisement blockers before performing the experiment so that MOPP's pop-up windows—used, for example, to provide progress feedback—can function properly, and to use the Chrome browser (MOPP was primarily tested on Chrome, and thus, there may be compatibility issues with other browsers).

**3.2. Authentication and verification**

MOPP is equipped with two tests to confirm participants' credibility:

(a) An *authentication test*. This is used to authenticate each participant's identity without collecting personal data. This test uses a third-party feature that enables participants to self-authenticate their identity by logging into their email accounts. A secure pop-out window is presented outside of MOPP's interface during this process. Successful login confirms the participant has a valid email address (those who do not have an active email account cannot pass this test). MOPP prevents multiple entries to the experiment by ensuring that once someone has accessed it, they cannot authenticate again using the same email account or IP address.

(b) A *verification test*. This is used to confirm that the participant is a human (and not an internet bot). For this, MOPP utilizes the *reCAPTCHA* test ((Von Ahn et al., 2008); the 2$^{nd}$ version of the Completely Automated Public Turing test to tell Computers and Humans Apart), which is embedded in MOPP's interface.

**3.3. Visual calibration**

MOPP incorporates two tests to improve experimental control for visual stimuli:

(a) The *virtual chinrest* test (Li et al., 2020). In this test, MOPP calculates the participant's viewing distance using trigonometry (based on the blind spot of the retina). The estimated viewing distance is then used to adjust the size of the subsequently presented stimuli to maintain a consistent visual angle of the stimulus on the retina. MOPP administers this test twice: once at the beginning and once at the end. These tests can be used for post-experimental analysis, correction, or data exclusion if the participant substantially changes their position during the experiment.

(b) The *Taylor's E* test (Bach, 2006). This is used to measure the participant's visual acuity. In this test, the participants are asked to indicate in which one of the four possible cardinal directions the open side of a rotated letter 'E' is pointing (e.g., without rotation, E is pointing rightward, reported by pressing the right keyboard arrow). The direction is randomly selected on each trial. Similar to previous studies (Bach, 2006), the task starts with four specific stimuli (the size of the E corresponds to an acuity of 0.1, 0.2, 0.4 and 0.8). The size of 'E' is then adapted individually according to each participant's performance, following a 2-up 1-down staircase procedure (Cornsweet, 1962). Namely, following two correct responses, the size of the E on next trial is reduced, and following one incorrect response, the size of the E on next trial is increased.

### 3.4. Perform experiment

Participants perform the tasks according to the order set by the researcher in advance (when building the experiment). MOPP displays the participant's progress in a progress bar with basic information regarding the experiment structure and duration (how many tasks and trials remain). During the tasks, MOPP checks that the correct data type was entered. Otherwise, it displays a "data type mismatch" message. At the end of the experiment, the researcher can provide a confirmation code on the completion page, which can be used by the participants to prove that they have completed the experiment. This is particularly useful for participants recruited via crowdsourcing platforms, where a completion code is required to receive compensation.

### 4. Download data

Once data collection is complete, the researcher can download the data from the experiment page (accessed via the researcher portal). The data is saved in a CSV file in a wide format - i.e., rows represent participants, and columns represent responses (with headers to specify the name of each variable). The participant's responses to the visual calibration test, and pre-experiment questions are also saved as columns.

## C. Pilot data results

We conducted a small online pilot study with MOPP (in public mode) to validate its functionality and to test whether the data collected would produce results consistent with those reported in laboratory settings. Seventeen participants (mean age ± SEM: 30.9 ± 2.6 years; 8 females) were recruited via online student forums. This study was approved by the internal review board at Bar-Ilan University (ethics committee approval number: ISU202110005), and all participants provided informed consent before participation. The pilot data and the code to generate the relevant figures are available at: https://github.com/YuvalSK/MOPP.

The mean viewing distance across participants, measured by the virtual chinrest test at the start and end of the experiment, was 54.1 ± 1.6 cm and 52.9 ± 1.9 cm, respectively (mean ± SEM: the range across both tests was 37.9 to 65.9 cm). This indicates participants were physically within an acceptable viewing distance from their PC screens. There was no significant difference in viewing distance between these two measurements ($p > 0.5$, $t(16) = 0.58$; $BF_{10} = 0.29$; two-tailed paired $t$-test). Thus, on average, there was no systematic shift during the experiment. The mean decimal visual acuity (VA), measured using Taylor's E test, was: 0.97 ± 0.01 (mean ± SEM). The VA of a person with a decimal VA value of 1 is considered normal, whereas values below 0.5 are generally considered indicative of a poor vision. Thus, on average, participants had close to normal vision. Six participants reported wearing glasses, and their visual acuity was similarly near normal (mean decimal VA = 0.99 ± 0.01).

### 1. Length task

In the length task, participants were asked to estimate, without counting, the length of a one-dimensional line. The line comprised a series of consecutive underbars. The participants were shown one underbar ('_') for reference, and then asked to estimate how many underbars a given stimulus contains (e.g., '__________' is 10 units; for elaboration on the task details, see Table 2). Stimulus magnitudes were randomly drawn from a Uniform distribution in the range of 1 to 18 (integers). This set of stimuli was generated once, and then used across all participants. Each participant completed 24 trials. Two participants were excluded due to missing trials (skipped more than one trial). This left 15 participants for further analysis in this task.

Responses to the length task are presented in Figure 3. Previous studies (Ekman & Junge, 1961; Stevens & Galanter, 1957) found that line length estimates are linearly related to the stimulus values. Accordingly, we fitted a linear regression to our data (Fig. 3, black regression line; $R^2$ = 0.93, $p$ < 0.001; fitting the mean values per stimulus). The regression line had a positive intercept ($\beta_0$ = 3.68; 95% CI = [2.70, 4.66]) and a slope smaller than 1 ($\beta_1$ = 0.51; 95% CI = [0.42, 0.60]) such that the fitted line crosses the unity line (at 7.51 units). This means that length stimuli on the lower end are overestimated, and those on the upper end are underestimated. This replicates the well-documented phenomenon of regression to the mean (Ashourian & Loewenstein, 2011; Petzschner et al., 2015).

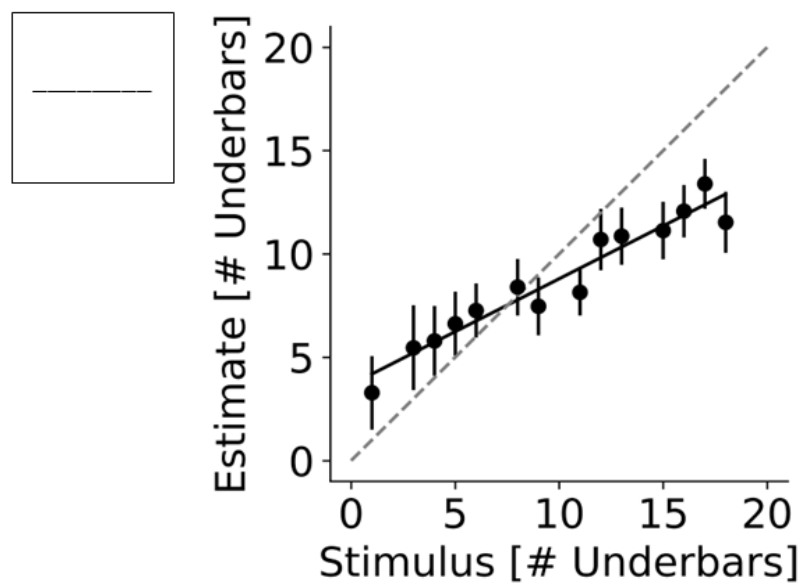

**Figure 3: Pilot data for the *length* task collected via MOPP.** A stimulus schematic is presented in the top left panel. In the main plot, black circles and error bars depict the mean ± SEM length estimates across participants (N = 15) as a function of the actual stimulus length. The solid black line depicts a linear regression of the group data, and the dashed gray line is the unity line (y = x). A set of 24 stimulus lengths was randomly drawn from a uniform distribution (integers in the range of 1-18) once, and then used for all participants.

## 2. Numerosity task

In the numerosity task, participants were asked to estimate, without counting, the number of dots in a two-dimensional array of black and white dots (Table 2). Stimulus magnitudes were randomly drawn from a Gaussian distribution (μ = 22, σ = 5 dots). This set of stimuli was generated once, and then used across all participants. Each participant completed 24 trials. No participants were excluded in this task (N = 17).

Responses to the numerosity task are presented in Figure 4. Previous studies found that numerosity estimates are systematically underestimated. Specifically, perceived numerosity increases with actual numerosity according to a power function, with an exponent smaller than one (Bevan & Turner, 1964; Burr & Ross, 2008; Crollen et al., 2013; Indow & Ida, 1977; Izard & Dehaene, 2008; Krueger, 1982). Namely, as the actual number of items increases, the perceived number increases more slowly—larger quantities are progressively underestimated relative to smaller ones.

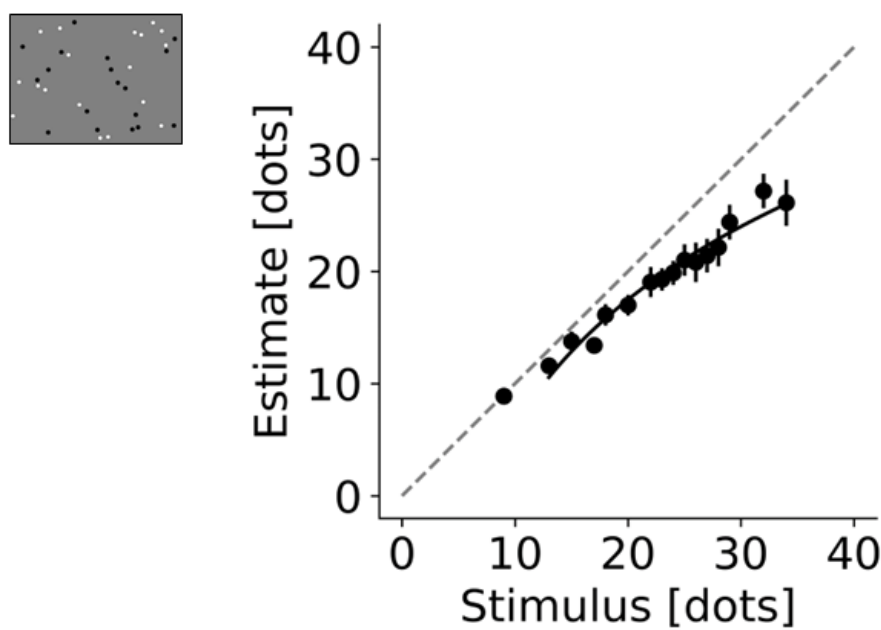

**Figure 4: Pilot data for the *numerosity* task collected via MOPP.** A stimulus schematic is presented in the top left panel. In the main plot, black circles and error bars depict the mean ± SEM numerosity estimates across participants (N = 17), as a function of the actual stimulus numerosity. The solid black line depicts a regression of the group data (fit in logscale), and the dashed gray line is the unity line (y = x). A set of 24 stimulus values (integers) was randomly drawn from a Gaussian distribution (mean = 22, SD = 5 dots) once, and then used for all participants.

To quantify this relationship in our data, we fitted a linear regression in log-log scale. By applying a log-transformation to both the stimulus magnitudes and participants' estimates, we could estimate the exponent of the power function from the slope of the regression line. The resulting fit, specifically the strong (Pearson's) correlation ($R^2$ = 0.97, p < 0.001; data fit only for stimuli exceeding ten dots) provides robust evidence for a linear relationship in the log-log scale. The regression line had an intercept of $\beta_0$ = 0.20 (95% CI = [-0.07, 0.47]) which did not differ significantly from zero, and a significantly positive slope of $\beta_1$ = 0.88 (95% CI = [0.79, 0.96]) which was also significantly less than 1. This indicates underestimation, namely, as the number of dots increases, the participants' estimates grew more slowly than the actual stimulus magnitude. Figure 4 presents the regression curve plotted on the original (non-logarithmic) scale for easier interpretability. This replicates the well-documented pattern of underestimation of numerosity.

## 3. Biological motion task

In the biological motion task, participants were presented with a set of ten white moving dots on a black background that reflected either biological motion or random motion (Table 2). They were asked to discriminate between the two. The same set of stimuli was used for all participants. Each participant completed 24 trials (half with biological motion and half with random motion). No participants were excluded in this task (N = 17).

Responses to the biological motion task are presented in Figure 5. For each participant, we calculated the proportion of trials for which they reported biological motion, for biological motion stimuli and random motion stimuli. We then calculated the mean of these proportions across participants. As expected, we observed higher rates of reporting biological motion for biological compared to random motion stimuli (Fig. 5; right vs. left bars, respectively; mean ± SEM = 93.0% ± 6.2% and 26.7% ± 10.7%, respectively; *p* < 0.001, *t*(16) = 10.1; one-tailed paired *t*-test).

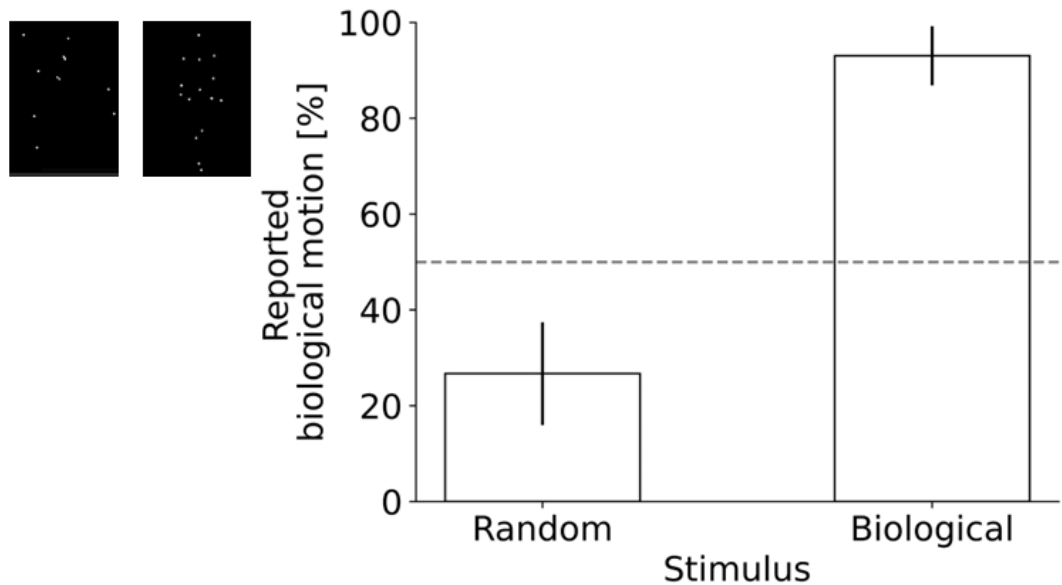

**Figure 5: Pilot data for the *biological motion* task collected via MOPP.** Stimulus schematics are presented on the top left (random motion and biological motion in the left and right schematics, respectively). The main plot presents the mean ± SEM reports of perceiving biological motion across participants (N = 17), per stimulus type. The dashed gray line marks chance level (50%). The same set of stimuli was used for all participants. The biological motion stimuli were generated using the tool developed by Troje (2002).

We used *d'* in order to measure individual sensitivity to biological motion detection, calculated as: *d'* = z(*H*) − z(*FA*), where *z* denotes the inverse of the cumulative standard normal distribution, *H* denotes the hit rate (reporting biological motion for biological stimuli), and *FA* denotes the false alarm rate (reporting biological motion for random stimuli). Higher *d'* values reflect better sensitivity. The *d'* in our data (mean ± SEM = 0.66 ± 0.07) did not differ significantly from the values reported by Weil et al (2018) for either their large online sample (0.73 ± 0.01, N = 189; *p* = 0.33, *t*(16) = -1.00; $BF_{10}$ = 0.40; two-tailed unpaired Welch's *t*-test) or their smaller offline sample (0.74 ± 0.03, N = 19; *p* = 0.30, *t*(16) = -1.06; $BF_{10}$ = 0.50; two-tailed unpaired Welch's *t*-test).

## 4. Mooney face task

In the Mooney face task, participants were presented with black-and-white images that either contained a face (oriented upright or inverted) or not (Table 2). They were asked to discriminate whether an image contains a face (regardless of orientation) or not (random image). The same set of stimuli was used for all participants. Each participant completed 24 trials (half trials without faces, and the other half equally split between upright and inverted faces). One participant who provided the same response on all trials was excluded. This resulted in 16 participants for further analysis in this task.

Results from the Mooney face task are presented in Figure 6. For each participant, we calculated the proportion of trials in which they identified a face for the different stimuli (upright faces, inverted faces and random images). We then calculated the mean of these proportions across participants. Consistent with the original study on which our stimuli were based (Schwiedrzik et al., 2018), we observed higher face detection rates for upright face images compared to inverted face images (Figure 6; right and center bars, respectively; mean ± SEM = 91.3% ± 1.8%, 30.5% ± 2.9%, respectively; $p < 0.001$, $t(30) = 7.92$; one-tailed unpaired $t$-test) and compared to random images (left bar; 17.0% ± 2.4%; upright vs. random: $p < 0.001$, $t(30) = 11.1$). A trend was observed of higher face detection rate for inverted images compared to random images ($p = 0.052$, $t(30) = 1.68$).

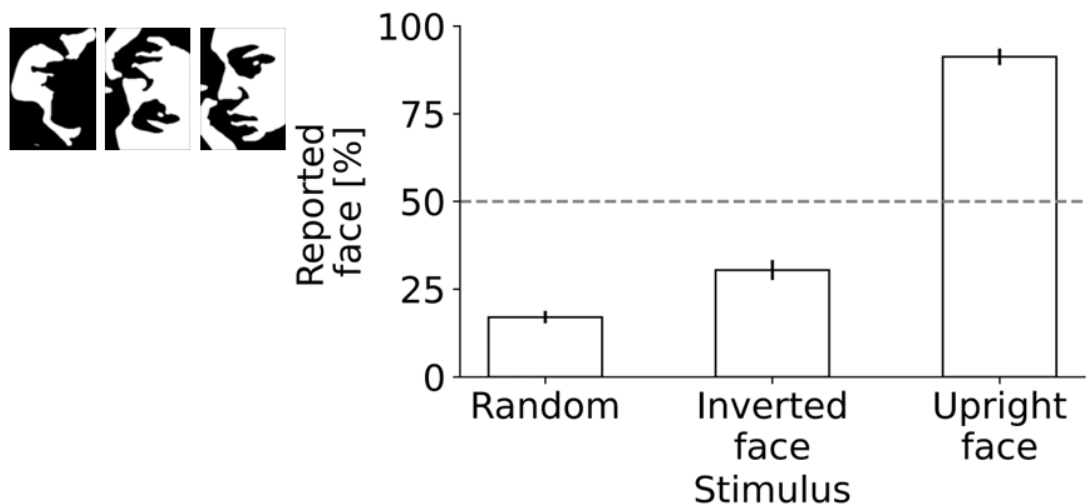

**Figure 6: Pilot data for the *Mooney face* task collected via MOPP.** Stimulus schematics are presented on the top left (from left to right: random, inverted, and upright faces). The main plot presents the mean ± SEM reports of perceiving faces across participants (N = 16), per stimulus type. The dashed gray line marks chance level (50%). The same set of stimuli was used for all participants. The Mooney images of faces were taken from the database made freely available by Schwiedrzik et al. (2018).

Here too, we used *d'* to measure individual sensitivity to face detection for each comparison. The overall *d'* in our data (mean ± SEM = 1.54 ± 0.65) did not differ significantly from the values reported by Schwiedrzik (2018) across trials and participants (1.19 ± 0.07, their N = 18; *p* = 0.65, *t*(16) = 0.53; $BF_{10}$ = 0.53; two-tailed unpaired Welch's *t*-test).

## 5. Key-tapping task

In the key-tapping task (Table 2), participants were asked to press the 's' and 'k' keys alternatingly on their keyboard as many times as they could within a 30-second interval using both hands, right hand only, or left hand only (three conditions). Each participant completed one trial per condition. Participants who failed to provide a response were excluded, per condition. This removed one participant in the both-hands condition and four participants in the left-hand condition. This resulted in 16, 17, and 13 participants for further analysis (for both hands, right-hand and left-hand conditions, respectively).

Responses to the key-tapping task are presented in Figure 7. The overall mean number of key taps using a single hand was 55.4 ± 7.0 taps (mean ± SEM; data pooled across right- and left-hand conditions). This was compared to the results from the original study on which the task was based (60.3 ± 1.4 taps, also using a single hand, with data pooled across right- and left-hand conditions; N = 93 participants; Noyce et al., 2014). No significant difference was observed between our data and the results from that original study (*p* = 0.50, t(31.3) = -0.69; two-tailed two-sample Welch's *t*-test; $BF_{10}$ = 0.27). The mean number of taps using both hands was 99.6 ± 11.0 (mean ± SEM). This condition could not be compared to previous studies, as prior research assessed key-tapping performance only for a single hand.

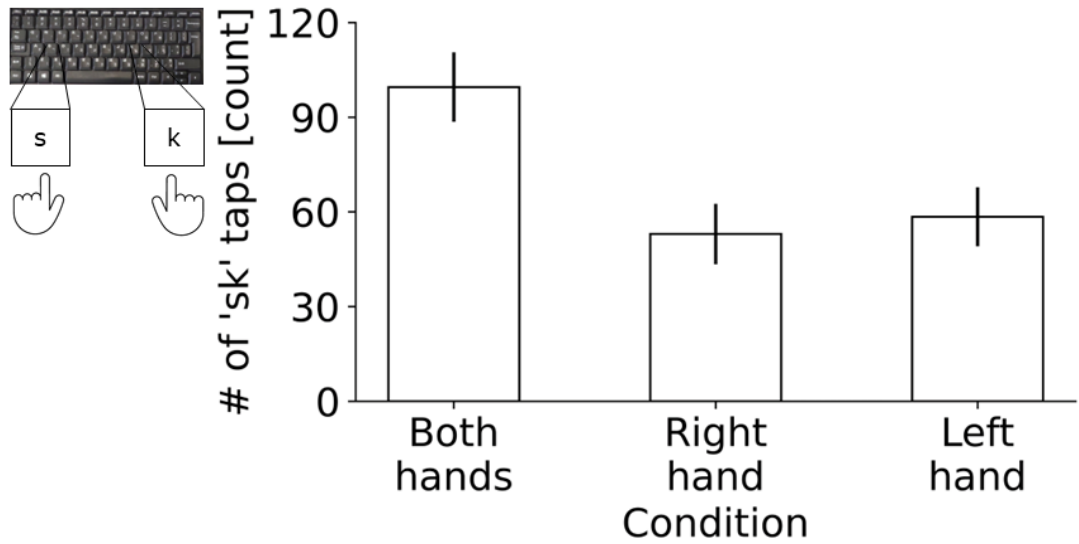

**Figure 7: Pilot data for the *key-tapping* motor task collected via MOPP.** An illustration of the task is presented in the schematic on the top left. Participants were required to sequentially press the 's' and 'k' buttons on their computer keyboard as many times as possible within a 30 s interval, in one of three conditions: using both hands, right hand only, or left hand only. The main plot presents the mean ± SEM number of taps across participants (N = 16, 17, and 13, respectively), per condition. Participants performed each condition once.

## Discussion

In this study, we introduce an open-source modular online psychophysics platform (*MOPP*) for researchers running online visual experiments. MOPP has a simple web-based interface to create modular experiments tailored to their scientific objectives. Additionally, it has integrated tools to confirm participants' credibility, calibrate for viewing distance, and measure visual acuity. This can enhance accessibility for researchers without a strong programming background, and facilitate comparison and replication of experiments.

To evaluate the data collected through MOPP, we developed five example psychophysics tasks that come preloaded in the environment and ran a pilot experiment online, hosted on the AWS cloud. In all five tasks, the data yielded results similar to previous publications collected in laboratory settings, validating our task implementations across the perceptual and motor domains.

## Limitations and future directions

Although it does not require programming knowledge, the installation process of MOPP currently comprises several steps. The MOPP guide, that details the process of hosting an experiment using the AWS cloud, aims to support and ease the installation process. A setup script is included in the guide to automate parts of the process, such as cloning the repository and configuring basic parameters. However, it does not have a stand-alone installer and requires several manual steps (e.g., cloud setup). Developing a cross-platform installation tool (e.g., via an execution file), perhaps by future users of this open-source platform, would help simplify this process and thereby further lower the technical barrier for researchers and students unfamiliar with cloud-based systems.

MOPP does not currently support mobile devices, as it is designed for traditional keyboard and mouse interactions. Besides its visual calibration tests, MOPP does not address several other inherent challenges of online experiments, such as variability in environmental conditions, monitors, and computer hardware. As a result, it cannot guarantee consistent and uniform stimulus presentation across devices (e.g., in terms of brightness or contrast). One possible mitigation strategy is to collect self-reported specifications – such as monitor model and other PC details - for post-experimental control and analysis. MOPP allows researchers to easily add such specifications questions in the pre-experiment questions phase without requiring additional coding. However, this approach relies on the participant's ability to provide this information accurately, and does not account for other environmental conditions (e.g., ambient light). Another solution is to compare online results to a controlled offline group, which can be run with MOPP's supervised mode.

MOPP is primarily designed to test visual psychophysics and does not currently support testing other sensory modalities (such as auditory, tactile etc.) or motor function (although some motor function can be assessed, e.g., via the keyboard tapping task, as we implemented). Researchers interested in studying these or multisensory processes would need to supplement MOPP with additional tools or devices. For instance, online auditory experiments might need auxiliary equipment, such as headphones with sound calibration tests to ensure reliable delivery of auditory cues (Milne et al., 2021; Schmack et al., 2021; Su et al., 2022). Experiments of more complex motor function, balance and spatial orientation, can use built-in sensors in smartphones, such as accelerometers and gyroscopes. These can also capture body and head movements (e.g., when placed in VR goggles) to assess vestibular function (Brodsky et al., 2015; Wengier et al., 2021).

## Concluding remarks

MOPP offers a simple web-based platform that allows researchers to create modular experiments tailored to their specific scientific goals. This enhances accessibility for researchers, including those without a strong programming background. MOPP can help researchers collect psychophysics datasets online, with reduced turnaround time, and in a standardized manner. By that, it complements traditional lab-based research and supports access to larger and more diverse populations outside the boundaries of a laboratory.

# References

Ashourian, P., & Loewenstein, Y. (2011). Bayesian inference underlies the contraction bias in delayed comparison tasks. *PLoS ONE*, *6*(5), 19551. https://doi.org/10.1371/journal.pone.0019551

Bach, M. (2006). The Freiburg Visual Acuity Test-Variability unchanged by post-hoc re-analysis. *Graefe's Archive for Clinical and Experimental Ophthalmology*, *245*(7), 965–971. https://doi.org/10.1007/S00417-006-0474-4,

Bevan, W., & Turner, E. D. (1964). Assimilation and contrast in the estimation of number. *Journal of Experimental Psychology*, *67*(5), 458–462. https://doi.org/10.1037/h0041141

Brainard, D. H. (1997). The Psychophysics Toolbox. *Spatial Vision*, *10*(4), 433–436. https://doi.org/10.1163/156856897X00357

Brodsky, J. R., Cusick, B. A., Kawai, K., Kenna, M., & Zhou, G. (2015). Peripheral vestibular loss detected in pediatric patients using a smartphone-based test of the subjective visual vertical. *International Journal of Pediatric Otorhinolaryngology*, *79*(12), 2094–2098. https://doi.org/10.1016/j.ijporl.2015.09.020

Burr, D., & Ross, J. (2008). A Visual Sense of Number. *Current Biology*, *18*(6), 425–428. https://doi.org/10.1016/j.cub.2008.02.052

Chandler, J., Mueller, P., & Paolacci, G. (2014). Nonnaïveté among Amazon Mechanical Turk workers: Consequences and solutions for behavioral researchers. *Behavior Research Methods*, *46*(1), 112–130. https://doi.org/10.3758/s13428-013-0365-7

Chmielewski, M., & Kucker, S. C. (2020). An MTurk Crisis? Shifts in Data Quality and the Impact on Study Results. *Social Psychological and Personality Science*, *11*(4), 464–473. https://doi.org/10.1177/1948550619875149

Cornsweet, T. N. (1962). The staircase method in psychophysics. *The American Journal of Psychology*, *75*(3), 485–491. https://doi.org/10.2307/1419876

Crollen, V., Grade, S., Pesenti, M., & Dormal, V. (2013). A common metric magnitude system for the perception and production of numerosity, length, and duration. *Frontiers in Psychology*, *4*(JUL), 449. https://doi.org/10.3389/fpsyg.2013.00449

de Leeuw, J. R. (2015). jsPsych: A JavaScript library for creating behavioral experiments in a Web browser. *Behavior Research Methods*, *47*(1), 1–12. https://doi.org/10.3758/s13428-014-0458-y

Druckman, J. N., & Kam, C. D. (2009). Students as Experimental Participants: A Defense of the "Narrow Data Base." *SSRN Electronic Journal*. https://doi.org/10.2139/SSRN.1498843

Ekman, G., & Junge, K. (1961). PSYCHOPHYSICAL RELATIONS IN VISUAL PERCEPTION OF LENGTH, AREA AND VOLUME. *Scandinavian Journal of Psychology*, *2*(1), 1–10. https://doi.org/10.1111/j.1467-9450.1961.tb01215.x

Grootswagers, T. (2020). A primer on running human behavioural experiments online. *Behavior Research Methods*, *52*(6), 2283–2286. https://doi.org/10.3758/s13428-020-01395-3

Indow, T., & Ida, M. (1977). Scaling of dot numerosity. *Perception & Psychophysics*, *22*(3), 265–276. https://doi.org/10.3758/BF03199689

Izard, V., & Dehaene, S. (2008). Calibrating the mental number line. *Cognition*, *106*(3), 1221–1247. https://doi.org/10.1016/J.COGNITION.2007.06.004

Kennedy, R., Clifford, S., Burleigh, T., Waggoner, P. D., Jewell, R., & Winter, N. J. G. (2020). The shape of and solutions to the MTurk quality crisis. *Political Science Research and Methods*, *8*(4), 614–629. https://doi.org/10.1017/psrm.2020.6

Krueger, L. E. (1982). Single judgments of numerosity. *Perception & Psychophysics*, *31*(2), 175–182. https://doi.org/10.3758/BF03206218

Lago, M. A. (2021). SimplePhy: An open-source tool for quick online perception experiments. *Behavior Research Methods*, *53*(4), 1669–1676. https://doi.org/10.3758/s13428-020-01515-z

Lange, K., Kühn, S., & Filevich, E. (2015). "Just Another Tool for Online Studies" (JATOS): An Easy Solution for Setup and Management of Web Servers Supporting Online Studies. *PLOS ONE*, *10*(6), e0130834. https://doi.org/10.1371/JOURNAL.PONE.0130834

Li, Q., Joo, S. J., Yeatman, J. D., & Reinecke, K. (2020). Controlling for Participants' Viewing Distance in Large-Scale, Psychophysical Online Experiments Using a Virtual Chinrest. *Scientific Reports*, *10*(1), 1–11. https://doi.org/10.1038/s41598-019-57204-1

Milne, A. E., Bianco, R., Poole, K. C., Zhao, S., Oxenham, A. J., Billig, A. J., & Chait, M. (2021). An online headphone screening test based on dichotic pitch. *Behavior Research Methods*, *53*(4), 1551–1562. https://doi.org/10.3758/s13428-020-01514-0

Noyce, A. J., Nagy, A., Acharya, S., Hadavi, S., Bestwick, J. P., Fearnley, J., Lees, A. J., & Giovannoni, G. (2014). Bradykinesia-akinesia incoordination test: Validating an online keyboard test of upper limb function. *PLoS ONE*, *9*(4). https://doi.org/10.1371/journal.pone.0096260

Onnela, J.-P., Dixon, C., Griffin, K., Jaenicke, T., Minowada, L., Esterkin, S., Siu, A., Zagorsky, J., & Jones, E. (2021). Beiwe: A data collection platform for high-throughput digital phenotyping. *Journal of Open Source Software*, *6*(68), 3417. https://doi.org/10.21105/joss.03417

Paolacci, G., & Chandler, J. (2014). Inside the Turk: Understanding Mechanical Turk as a Participant Pool. *Current Directions in Psychological Science*, *23*(3), 184–188. https://doi.org/10.1177/0963721414531598

Peer, E., Rothschild, D., Gordon, A., & Damer, E. (2022). Erratum to Peer et al. (2021) Data quality of platforms and panels for online behavioral research. *Behavior Research Methods*, *54*(5), 2618–2620. https://doi.org/10.3758/s13428-022-01909-1

Peirce, J., Gray, J. R., Simpson, S., MacAskill, M., Höchenberger, R., Sogo, H., Kastman, E., & Lindeløv, J. K. (2019). PsychoPy2: Experiments in behavior made easy. *Behavior Research Methods*, *51*(1), 195–203. https://doi.org/10.3758/s13428-018-01193-y

Petzschner, F. H., Glasauer, S., & Stephan, K. E. (2015). A Bayesian perspective on magnitude estimation. *Trends in Cognitive Sciences*, *19*(5), 285–293. https://doi.org/10.1016/j.tics.2015.03.002

Rajananda, S., Lau, H., & Odegaard, B. (2018). A random-dot kinematogram for web-based vision research. *Journal of Open Research Software*, *6*(1). https://doi.org/10.5334/jors.194

Schmack, K., Bosc, M., Ott, T., Sturgill, J. F., & Kepecs, A. (2021). Striatal dopamine mediates hallucination-like perception in mice. *Science*, *372*(6537). https://doi.org/10.1126/science.abf4740

Schwarzbach, J. (2011). A simple framework (ASF) for behavioral and neuroimaging experiments based on the psychophysics toolbox for MATLAB. *Behavior Research Methods*, *43*(4), 1194–1201. https://doi.org/10.3758/s13428-011-0106-8

Schwiedrzik, C. M., Melloni, L., & Schurger, A. (2018). Mooney face stimuli for visual perception research. *PLoS ONE*, *13*(7), e0200106. https://doi.org/10.1371/journal.pone.0200106

Sears, D. O. (1986). College Sophomores in the Laboratory. Influences of a Narrow Data Base on Social Psychology's View of Human Nature. *Journal of Personality and Social Psychology*, *51*(3), 515–530. https://doi.org/10.1037/0022-3514.51.3.515

Stevens, S. S., & Galanter, E. H. (1957). Ratio scales and category scales for a dozen perceptual continua. *Journal of Experimental Psychology*, *54*(6), 377–411. https://doi.org/10.1037/h0043680

Su, Z. H., Patel, S., Bredemeyer, O., FitzGerald, J. J., & Antoniades, C. A. (2022). Parkinson's disease deficits in time perception to auditory as well as visual stimuli – A large online study. *Frontiers in Neuroscience*, *16*. https://doi.org/10.3389/fnins.2022.995438

Torous, J., Kiang, M. V., Lorme, J., & Onnela, J. P. (2016). New tools for new research in psychiatry: A scalable and customizable platform to empower data driven smartphone research. *JMIR Mental Health*, *3*(2), e5165. https://doi.org/10.2196/mental.5165

Troje, N. F. (2002). Decomposing biological motion: A framework for analysis and synthesis of human gait patterns. *Journal of Vision*, *2*(5), 371–387. https://doi.org/10.1167/2.5.2

Von Ahn, L., Maurer, B., McMillen, C., Abraham, D., & Blum, M. (2008). reCAPTCHA: Human-based character recognition via web security measures. *Science*, *321*(5895), 1465–1468. https://doi.org/10.1126/science.1160379

Weil, R. S., Schwarzkopf, D. S., Bahrami, B., Fleming, S. M., Jackson, B. M., Goch, T. J. C., Saygin, A. P., Miller, L. E., Pappa, K., Pavisic, I., Schade, R. N., Noyce, A. J., Crutch, S. J., O'Keeffe, A. G., Schrag, A. E., & Morris, H. R. (2018). Assessing cognitive dysfunction in Parkinson's disease: An online tool to detect visuo-perceptual deficits. *Movement Disorders*, *33*(4), 544–553. https://doi.org/10.1002/mds.27311

Wengier, A., Ungar, O. J., Handzel, O., Cavel, O., & Oron, Y. (2021). Subjective Visual Vertical Evaluation by a Smartphone-based Test - Taking the Phone out of the Bucket. *Otology and Neurotology*, *42*(3), 455–460. https://doi.org/10.1097/MAO.0000000000002944

Woods, A. T., Velasco, C., Levitan, C. A., Wan, X., & Spence, C. (2015). Conducting perception research over the internet: A tutorial review. *PeerJ*, *2015*(7). https://doi.org/10.7717/peerj.1058